# Deep learning-enhanced paper-based vertical flow assay for high-sensitivity troponin detection using nanoparticle amplification

Gyeo-Re Han[1], Artem Goncharov[1], Merve Eryilmaz[1,2], Hyou-Arm Joung[1], Rajesh Ghosh[2], Geon Yim[3], Nicole Chang[2], Minsoo Kim[2], Kevin Ngo[2], Marcell Veszpremi[1], Kun Liao[1], Omai B. Garner[4], Dino Di Carlo[2,5*], and Aydogan Ozcan[1,2,5,6*]

[1]Electrical & Computer Engineering Department, [2]Bioengineering Department, [3]Department of Chemistry and Biochemistry, [4]Department of Pathology and Laboratory Medicine, [5]California NanoSystems Institute (CNSI), [6]Department of Surgery, University of California, Los Angeles, CA 90095 USA.

*Corresponding Authors: dicarlo@ucla.edu, ozcan@ucla.edu

**Abstract:**

Successful integration of point-of-care testing (POCT) into clinical settings requires improved assay sensitivity and precision to match laboratory standards. Here, we show how innovations in amplified biosensing, imaging, and data processing, coupled with deep learning, can help improve POCT. To demonstrate the performance of our approach, we present a rapid and cost-effective paper-based high-sensitivity vertical flow assay (hs-VFA) for quantitative measurement of cardiac troponin I (cTnI), a biomarker widely used for measuring acute cardiac damage and assessing cardiovascular risk. The hs-VFA includes a colorimetric paper-based sensor, a portable reader with time-lapse imaging, and computational algorithms for digital assay validation and outlier detection. Operating at the level of a rapid at-home test, the hs-VFA enabled the accurate quantification of cTnI using 50 µL of serum within 15 min per test and achieved a detection limit of 0.2 pg/mL, enabled by gold ion amplification chemistry and time-lapse imaging. It also achieved high precision with a coefficient of variation of < 7% and a very large dynamic range, covering cTnI concentrations over six orders of magnitude, up to 100 ng/mL, satisfying clinical requirements. In blinded testing, this computational hs-VFA platform accurately quantified cTnI levels in patient samples and showed a strong correlation with the ground truth values obtained by a benchtop clinical analyzer. This nanoparticle amplification-based computational hs-VFA platform can democratize access to high-sensitivity point-of-care diagnostics and provide a cost-effective alternative to laboratory-based biomarker testing.

## Introduction

Cardiovascular diseases (CVDs) present a major public health challenge worldwide, representing a leading cause of mortality, with approximately 19.1 million deaths in 2020, accounting for roughly 32% of global deaths, and imposing a significant financial burden on patients and healthcare systems[1,2]. From a socioeconomic standpoint, the incidence of CVDs shows an inverse correlation with income, with more than 80% of CVD-related deaths globally reported in low- and middle-income countries, exacerbating the global burden of CVDs shifting to these regions[3]. Similarly, within the U.S., the burden of CVDs is increased in lower-income communities[4,5]. Amid the spectrum of CVDs, acute myocardial infarction (AMI), commonly referred to as a heart attack, is a particularly critical event characterized by a sudden and severe impairment of blood flow due to a blockage of the coronary artery[6]. Chest pain that may indicate AMI is a primary reason for emergency department visits, with suspected AMI patients accounting for ~10% of all emergency department visits[7]. Furthermore, AMI stands as a paramount contributor to sudden death, with its swift onset and severe consequences, making it a primary concern in cardiovascular emergencies[8].

To date, numerous clinical care advancements related to AMI have benefitted from laboratory-based high-sensitivity (hs) cardiac troponin I (cTnI) assays[9,10]. cTnI is a protein specific to myocardial cells and is released into the bloodstream in the case of myocardial damage, making it the gold standard biomarker for the diagnosis of AMI[5]. The latest clinical guidelines, based on the fourth universal definition of myocardial infarction[11], prioritized the identification of myocardial injury by small increases in cTnI levels as an essential measure in diagnostic algorithms along with other evidence (e.g., electrocardiography or imaging), thereby reinforcing the role of cTnI measurement in the diagnosis of AMI. Currently available, clinically deployed benchtop hs-cTnI assays can quantify cTnI at low concentrations (typically a few pg/mL) at or below the 99$^{th}$ percentile upper reference limit (URL) or in more than 50% of the healthy population with a high precision (coefficient of variation [CV] ≤ 10%)[12], providing opportunities for early diagnosis, rapid rule-in/rule-out, risk categorization, and prognostic applications[9,10,13]. However, laboratory-based hs-cTnI assays and readout instruments are relatively expensive and bulky, and require specialized/trained operators. As a result, they are mainly available in large hospitals with well-developed medical infrastructure, making these assays inaccessible to remote and low-income regions. Underserved populations in these regions have a higher prevalence of AMI[4,5], and the need for rapid and accurate diagnosis is critical[7,14,15]. Even in cases where a patient suspected of AMI reaches the hospital within the golden hour (the first hour after the onset of symptoms), the medical protocol conventionally followed is complicated and multi-step[16] (i.e., ordering a test from a healthcare provider, collecting and transporting a specimen to a central facility, processing the specimen, and reporting the results; see **Fig. 1a**). Turnaround times for this procedure can take over several hours[17], further compounded by a 2–3 h waiting period that patients with intermediate cTnI levels (between normal and abnormal ranges) must undergo for follow-up testing. This follow-up testing is necessary to monitor the potential elevation of cTnI levels and make a conclusive determination of safe rule-in or rule-out[18,19]. Moreover, high-volume sample processing and batch testing done at clinical laboratories typically cause additional delays in reporting results and performing critical medical interventions[20], making this framework not ideal for addressing the time-sensitive nature of AMI, where rapid diagnosis and response are essential.

Coupled with the continuous advances in sensing, information and communication technologies, point-of-care testing (POCT) is emerging as a viable alternative approach in the diagnostic landscape of AMI, potentially transforming diagnostic pipelines by delivering rapid, affordable, and on-site diagnostic measures (**Fig. 1b**)[7,16,17,21]. POCT technologies for AMI can provide the advantage of decentralized testing, which expands testing capabilities in various healthcare settings (e.g., hospitals, clinics, nursing homes, pharmacies) and accessibility even in resource-limited areas. By circumventing the complexities and delays associated with the traditional workflow, the POCT of cTnI could expedite clinical actions to be performed within 20 min and reduce the time and cost linked to the waiting period before the follow-up testing. To address this important need, several cTnI-POCT products are currently available[7,10,22], including those based on microfluidics[23-27], lateral flow assays[28], and automated enzyme-linked immunosorbent assays[29,30]. Some of the additional recent advances also involve microarray technologies[31], paper-based sensors[32-34], electrode-based sensors[35-37], and surface-enhanced Raman scattering[38] coupled with luminescent, electrochemical, and colorimetric sensing modalities. However, it is important to note that because of the extremely low clinical cut-off level of cTnI compared to other biomarkers, most of these POCT assays currently have limitations, particularly in terms of their sensitivity and precision (see **Fig. 1c** and **Table 1**)[7,11]. Additionally, some of these assays require expensive benchtop hardware and rely on AC power[5,29,30], which limits their portable use. Even though assay platforms can be miniaturized with hand-held analyzers[23,25], the high hardware costs exceed $15K and create another barrier to widespread implementation in clinical settings, particularly in low- and middle-income countries. These necessitate continued research and development to create high-performance cTnI-POCT solutions that meet clinical requirements for high sensitivity and precision while ensuring cost-effectiveness.

Here, we report a paper-based high-sensitivity vertical flow assay (hs-VFA) for rapid and cost-effective POCT of cTnI, providing enhanced sensitivity and high precision, and meeting the clinical standards for hs-cTnI testing. This hs-VFA system consists of a paper-based sensor coupled with a rapid nanoparticle amplification chemistry, a portable reader with time-lapse imaging capability, and computational analysis algorithms empowered by deep learning (**Fig. 1d**). This system enabled highly sensitive and accurate quantification of cTnI from 50 µL of serum within 15 min per test. We developed a paper-based sensor and assay cartridge to perform the immunoassay and signal amplification with simple operational steps, comparable to at-home rapid diagnostic tests. We validated the performance of our hs-VFA by testing it using cTnI spiked serum samples as well as clinical patient serum samples collected from healthy and diseased donors. Signals from the activated immunoassays were captured by a custom-designed, cost-effective and hand-held reader in a time-lapse manner. The acquired data were processed by computational assay validation and outlier analysis algorithms to improve assay performance, resulting in a limit of detection (LoD) of 0.2 pg/mL with high precision (average CV < 7%). Furthermore, hs-VFA signals were used to train neural network-based models that accurately quantified cTnI levels in clinical serum samples obtained from patients. When blindly tested on a set of 46 serum samples from 23 patients, the predictions of the computational hs-VFA showed a high correlation (Pearson's $r = 0.965$) with the ground truth measurements obtained by an FDA-cleared benchtop analyzer, while achieving ~93.5% accuracy for classifying the concentration of the serum samples to be below/above the cut-off of the benchtop clinical analyzer (< 40 pg/mL), which increased to 100% accuracy using duplicate testing per patient. This study demonstrates a significant leap forward in hs-VFA technology for POCT and offers an affordable, robust, and high-performance sensing platform for cTnI and potentially other critical low-abundance biomarkers used for rapid diagnoses and medical interventions.

## Results
### Design and optimization of the hs-VFA platform
The hs-VFA was engineered to achieve highly sensitive biomarker detection while maintaining the essential features of conventional VFAs[39-42], such as cost-effectiveness, simplicity, user-friendliness, and rapid assay time. The hs-VFA cartridges comprise a bottom case with a capture antibody-spotted sensing membrane, along with two top cases, each serving distinct roles of the immunoassay (1st top) and the signal amplification chemistry (2nd top) (**Fig. 2a**). Wax printing created multiple hydrophilic compartments on the sensing membrane, surrounded by hydrophobic wax layers, thereby concentrating fluid flow into the reaction areas. Among the 17 reaction regions of the hs-VFA cartridge, 10 were assigned as testing spots (treated with cTnI capture antibodies), 2 functioned as positive control spots (treated with secondary antibodies that bind the detection antibodies), and the remaining 5 were designated as negative control spots (blanks). This arrangement allowed for 10 repeated tests within a single assay, which was important to mitigate test-to-test variations and flow non-uniformity-related imperfections usually observed in inexpensive POCT sensors.

Our hs-VFA utilized passive capillary flow for vertical sample transport. The 1$^{st}$ top case facilitated uniform three-dimensional fluid flow, ensuring an even distribution of small-volume samples (≤ 50 µL) and reagents across open compartments on the sensing membrane (12 × 12 mm$^2$). The 2$^{nd}$ top case was fabricated by assembling a 3D-printed cartridge with a transparent acrylic cover and a gasket. An amplification reagent solution, introduced through the inlet, flowed directly onto the sensing membrane via the enclosed chamber, which served as a reservoir. The reagent solution evenly spread across the entire membrane in < 1 s, enabling uniform reaction across all the test spots.

A custom-designed portable hs-VFA reader was assembled using a Raspberry Pi computer, a touch display, and cost-effective optical components (i.e., camera module, macro lens, and light-emitting diodes [LEDs], see the "Portable reader and image analysis" subsection in the Methods). This configuration enabled high-resolution imaging of the sensing membrane within a compact, portable design ideal for POCT applications (**Fig. 2b**). The cassette tray blocked external ambient light and connected to the main housing through a sliding mechanism. Time-lapse images of the sensing membrane (for each test) were taken under 532 nm LED illumination. The graphical user interface enabled real-time observations of the sensing membrane image and provided adjustable imaging parameters (i.e., exposure time, total number of images, and time interval) for optimal time-lapse imaging (**Fig. S1**).

The entire assay process runs within 15 min per test, encompassing two successive phases of the assay (immunoassay and signal amplification) followed by the signal readout (**Fig. 2c–d**). For the immunoassay using the 1$^{st}$ top case, the cartridge is initially filled with a running buffer to activate the immobilized capture antibodies on the sensing membrane. Then, a mixture of the serum sample (to be tested) and 15 nm gold nanoparticle (AuNP)-detection antibody conjugates is introduced, allowing the recognition of antigens. As this mixture flows through the sensing membrane, a secondary antigen recognition event occurs upon binding to capture antibodies, resulting in the accumulation of the detection antibody conjugate as a function of antigen levels. Finally, a running buffer is injected to expedite the downward flow of the sample-conjugate mixture toward the absorption pads, accelerating the washing process in the stacked papers.

After the 10-min immunoassay step, the 1$^{st}$ top case is replaced with the 2$^{nd}$ top case, and the amplification solution (500 µL) containing gold ions (Au$^{3+}$) and a reducing agent (H$_3$NO) is introduced to initiate the signal amplification. This process amplifies the colorimetric response of each reaction area where AuNPs were previously bound. Notably, the AuNPs catalyze the reduction of the Au$^{3+}$ on their surfaces in the presence of H$_3$NO, leading to gold particle growth on the paper (**Fig. 2d**). This results in a more intense optical signal per test spot, primarily due to increased absorption, further enhancing the assay's sensitivity to <1 pg/mL. **Fig. 2e** shows brightfield images of the sensing membrane alongside the corresponding scanning electron microscope (SEM) images of the testing spots after the immunoassay and signal amplification stages. These images demonstrate a substantial increase in the original AuNP diameter (from 15 nm to 100–200 nm) and the formation of particle clusters following the signal amplification, providing direct evidence of the physical changes occurring during the Au-ion reduction strategy for assay signal amplification. Furthermore, **Video S1** presents an example of the real-time colorimetric response of the signal amplification reaction. The flow mechanism of the amplification reagent solution (**Fig. S2**) and spectroscopic analysis of the amplification reaction (**Fig. S3**) are also provided in Supplementary Note 1.

We assessed the uniformity of this signal amplification step across all the reaction spots, revealing a decrease in CV, from 7.9% to 4.1%, when excluding the edge spots (**Fig. 2f**). Regarding the analytical sensitivity of the VFA (**Fig. 2g**), we experimentally verified a 212.3-fold improvement in sensitivity threshold due to the Au-ion reduction, defined as $Mean\ blank\ value + 3 \times Standard\ deviation\ (SD)\ of\ the\ blank\ value$, compared to that of the original AuNP conjugate (0.0009 optical density [OD] vs. 0.1911 OD, respectively). This major signal enhancement elevates the sensitivity of the hs-VFA platform to the sub pg/mL range, a major improvement over conventional VFAs that typically operate at ng/mL levels of LoD[39,40]. Further analyses on sensor performance optimization for improved signal-to-noise ratio, cartridge compressibility (**Fig. S4a**), AuNP size optimization (**Fig. S4b**), running buffer combinations (**Fig. S4c**), and signal amplification reagent concentrations (**Fig. S5**) are provided in Supplementary Note 2.

**Time-lapse imaging using a hand-held quantitative reader**

Time-lapse imaging involves the periodic capture of images over a specified time period, which can play a pivotal role in understanding the dynamic nature of bio/chemical reactions, analyzing assay kinetics, and reducing the overall measurement and sensing time[43,44]. We utilized time-lapse imaging to improve the sensitivity and precision of our hs-VFA. Coupled with the signal amplification strategy, we periodically imaged the ongoing Au-ion reduction process. We hypothesized that this approach would mitigate some of the non-specific signals typically associated with prolonged reaction times in conventional end-point measurements, thus allowing for more sensitive measurements.

To test this hypothesis, we compared the time-lapse and end-point imaging methods using the cTnI assay. In the time-lapse approach, we restricted data acquisition to three images sampled at 30s intervals (t = 0, 30, 60s) to avoid the generation of excessive data with even shorter capture intervals (**Fig. 2h**). We then computed the time-lapse hs-VFA signal obtained from these images to measure the intensity increase over time (see the "Limit of detection of cTnI hs-VFA" subsection for the definition of the time-lapse hs-VFA signal). In the end-point measurement method, however, the intensity was measured from a single image captured at t = 360s when the signal intensity saturated (**Fig. 2h**). An end-point at t = 60s provided insufficient signal intensity, preventing the differentiation of low cTnI levels with acceptable statistical significance ($P < 0.05$). For comparison, we normalized each dataset by dividing the intensity values using the negative control results. Our analyses showed that the time-lapse imaging approach significantly outperformed the end-point measurement and enhanced the assay sensitivity by 6.5-fold, also reducing the CV in low cTnI levels ($\leq 10$ pg/mL) from 11.0% to 4.2% and shortening the readout time to 1 min (**Fig. 2i**). Based on these analyses, we adopted time-lapse imaging (with t = 0, 30, 60s) as part of our hs-VFA concentration measurement strategy, which will be further detailed in the following subsections.

**Computational data processing: virtual control indicator and outlier analysis**

Incorporating assay quality control algorithms ensures the precision and reliability of our hs-VFA platform. To achieve this, we developed a two-tier computational data processing approach, which included a virtual control indicator (VCI) and outlier analysis (OA) protocol (**Fig. 3a**). We hypothesized that the incorporation of the VCI would act as an internal benchmark to monitor assay failure, and the OA would enable the identification and exclusion of outlier test data points per test. By strategically leveraging these digital data processing techniques, we improved the robustness, sensitivity, and precision of the hs-VFA results.

The VCI functions as the first algorithmic step that assesses the validity of the hs-VFA results (**Fig. 3a, inset i**). In contrast to conventional immunoassays, like lateral flow assays that rely on binary on/off responses from the control area for assay validation, we employed a more advanced approach. We constructed a VCI max/min threshold by consolidating data from 120 hs-VFA test results. To establish the max/min threshold for determining assay validity, we used ±1.96 SD from the mean of the intensity ratio between the positive (i.e., $I_{Pos}^{j,t}$) and the negative (i.e., $I_{Neg}^{j,t}$) control spots; see the Methods subsection "Computational analysis of hs-VFA signals". This VCI signal range is effectively used to filter out any false signals that could occur due to interference or non-specific interactions during the hs-VFA operation. If anomalies arise due to assay failure or sample-related issues, the VCI signal will be out of its acceptable range, and the system will display an "invalid" message, suggesting the user perform further dilution and/or retesting. The impact and details of the application of VCI on the processing of clinical samples are detailed in the "Limit of detection of cTnI hs-VFA" subsection of the Results.

Following the VCI-based virtual quality control that utilizes signals from the positive and negative control spots, the second digital step, OA, is used to statistically validate the quality of the assay data using the raw intensities of the 10 test spots (i.e., $I_{Test}^{j,t}$). As shown in the **inset ii of Fig. 3a**, the test data points that lie outside of the 95% confidence interval (CI) level are excluded. This process helps mitigate spot-to-spot variations that may be introduced by potential errors, such as nonuniformity in the fluid flow within the paper matrix, non-specific binding, or irregularities in antibody spotting. As shown in **Fig. 3b**, the OA algorithm significantly improves intra-assay uniformity (from a CV of 4.1% to a CV of 1.1%) and inter-assay reproducibility (from a CV of 8.2% to a CV of 5.1%). Additional details on the design and performance of this OA algorithm are provided in Supplementary Note 3. The impact of the OA algorithm on improving assay sensitivity is detailed in the next subsection.

**Limit of detection of cTnI hs-VFA**

The LoD of the optimized hs-VFA was evaluated using titration experiments with different concentrations of cTnI spiked and serially diluted in cTnI-free human serum. The signal intensity of the optimized hs-VFA with signal amplification, time-lapse imaging, and OA was calculated according to equation (1):

$$I_{Time-lapse} = \bar{I}_{Test,Ref}^2 - \bar{I}_{Test,Ref}^1 + \bar{I}_{Test,Ref}^3 - \bar{I}_{Test,Ref}^1, \qquad (1)$$

where $\bar{I}_{Test,Ref}^t = I_{Test,Ref}^t - I_{Neg,Ref}^t$ is the OA-refined test signal at time point $t$ ($t$=1,2,3 - corresponding to 0, 30, 60s, respectively) after subtraction of the negative spot signal (refer to "Computational analysis of hs-VFA signals" subsection in Methods for the definition of $I_{Test,Ref}^t$ and $I_{Neg,Ref}^t$). Normalized time-lapse signal ($I_{Normalized,\ Time-lapse}$; **Fig. 3c–d**) was obtained as:

$$I_{Normalized,\ Time-lapse} = \frac{I_{Time-lapse}}{I_{Time-lapse}^{Neg\ samp}}, \qquad (2)$$

where $I_{Time-lapse}^{Neg\ samp}$ is the time-lapse hs-VFA signal for the negative control sample (i.e., 0 pg/mL cTnI).

The results of the titration assay revealed that the hs-VFA provided a response that was proportional to varying cTnI concentrations across the clinically relevant range ($10^0$–$10^5$ pg/mL)[18] (**Fig. 3c** and **Fig. S6**). The detectable range spanned *six orders of magnitude* without signal degradation due to the hook effect. Signal saturation was observed above the cTnI level of $10^5$ pg/mL, beyond the clinically relevant range. Based on these measurements, the LoD of the hs-VFA was determined to be 0.2 pg/mL using the following equations[31,45]:

$$LoD = Limit\ of\ Blank\ (LoB) + 1.645 \times SD\ of\ the\ lowest\ measured\ cTnI\ value, \qquad (3)$$

$$LoB = Mean\ blank\ value + 1.645 \times SD\ of\ the\ blank\ value, \qquad (4)$$

where the time-lapse hs-VFA signal ($I_{Normalized,\ Time-lapse}$) of the *Mean blank value*, *SD of the blank value*, and the *SD of the lowest measured cTnI value* were 1.000, 0.029, and 0.103 pg/mL, respectively; and the lowest measured cTnI value was 1 pg/mL. The LoB value was further estimated at 0.13 pg/mL by converting the time-lapse hs-VFA signal ($I_{Normalized,\ Time-lapse}$, y-value) to a concentration value (x-value) using the optimal calibration curve in **Fig. 3c** (i.e., with OA, $y = 2.2x^{0.36}$). Finally, the LoD value was calculated as 0.2 pg/mL using the same optimal fitting method.

In addition, as demonstrated in **Fig. 3c**, the assay results using OA showed very good precision (with a CV of 3.0 ± 1.6%) and a coefficient of determination ($R^2$) of 0.998. However, without OA, a relatively worse performance was reported, with an LoD of 0.42 pg/mL, a CV of 6.1 ± 4.3%, and an $R^2$ of 0.986, confirming the impact of OA in improving the precision and robustness of the hs-VFA.

The statistical validation of the assay results for low cTnI levels at or below the 99[th] percentile URL level (1–50 pg/mL) showed that all the *P*-values were lower than 0.001 (**Fig. 3d**). This demonstrates that the hs-VFA can effectively differentiate between cTnI concentrations, even at levels below the clinical cut-off.

These results and the overall performance of the hs-VFA adhere to two crucial criteria outlined in clinically recommended guidelines for hs-cTnI testing[12]: (i) achieving a CV of ≤10% at the 99[th] percentile URL, and (ii) detecting concentrations at or above the assay's LoD in over 50% of healthy individuals (i.e., a few pg/mL to 50 pg/mL), demonstrating our approach's alignment with the well-established clinical standards.

Next, we performed clinical sample testing using 62 samples, including 54 patient serum samples obtained from UCLA Health and 8 derived samples produced by diluting two of these patient samples with cTnI-free serum. The ground truth values were determined using an FDA-cleared benchtop analyzer, with cTnI concentrations below 40 pg/mL recorded as "< 40 pg/mL" due to the equipment's cut-off level. The assay response plot demonstrated an increasing signal with rising cTnI concentrations (**Fig. 3e, inset i**); see the "Computational analysis of hs-VFA signals" subsection of the Methods. Four samples with cTnI concentrations ≥ 40 pg/mL showed significant deviations from the expected trend. However, the application of VCI and OA successfully identified and excluded these four outlier samples (**Fig. 3e, inset ii**), leading to enhanced precision (average CV reduced from 5.4% to 2.7% after OA). The variations observed between the time-lapse assay signals (i.e., $I_{w/o\ Normalized,\ Time-lapse}$) and the ground truth measurement values may be due to (i) potential troponin degradation during sample storage[46], (ii) biochemical interferents such as lipemia or elevated bilirubin[47], or (iii) the presence of cTnI autoantibodies or heterophile antibodies in patient samples[47]. **Table S1** provides detailed results corresponding to the cases before and after applying VCI and OA for each clinical serum sample.

To infer cTnI concentrations from the captured hs-VFA signals, we next applied trained neural network models for accurate quantification of cTnI in serum samples. The impact of deep learning-based cTnI quantification based on the hs-VFA signals will be detailed in the following subsection.

**Neural network-based analysis of cTnI hs-VFA and blind testing with clinical samples**

Neural networks consist of multiple interconnected layers with non-linear activation functions, which allow for robust quantitative analysis of biological samples with point-of-care (POC) sensors despite additional noise from the sample matrix and the low-cost nature of these sensors[39-42]. In this work, we used fully connected neural networks to measure cTnI concentrations in clinical serum samples from the time-lapse hs-VFA signals (i.e., $I_{Time-lapse}$, defined by equation (1)). Our neural network-based hs-VFA analysis consisted of two successive parts: (i) *classification* of serum samples as either ≥ 40 pg/mL or < 40 pg/mL since the clinical cut-off of the ground truth benchtop cTnI analyzer used was 40 pg/mL; and (ii) *quantification* of cTnI concentration for the samples in the ≥ 40 pg/mL range (**Fig. 4a**). For this cTnI measurement/inference task, we first optimized the architecture of the neural network models using a portion of the samples (validation set). Based on this optimization (detailed in the Methods section), we converged to 3 individual fully-connected neural networks that collaborate with each other as illustrated in **Fig. 4**: $DNN_{Classification}$ neural network is used for the classification between the two concentration ranges (i.e., ≥ 40 pg/mL or < 40 pg/mL; **Fig. S7**) and the other two neural networks are used for cTnI quantification (i.e., $DNN_{Quantification}$ and $DNN_{Low}$; **Fig. S8**). If $DNN_{Classification}$ revealed a classification decision of < 40 pg/mL, that was the final decision, and the quantification neural networks ($DNN_{Quantification}$ or $DNN_{Low}$) were *not* used. The quantification stage was only used when $DNN_{Classification}$ blindly classified the sample as ≥ 40 pg/mL. In this quantification stage, the serum sample measurement was first processed by the $DNN_{Quantification}$ model and its inference revealed the cTnI concentration of the serum sample; this blind inference of $DNN_{Quantification}$ was used as our final cTnI concentration measurement if it was ≥ 40 pg/mL - i.e., complying with the former inference of $DNN_{Classification}$ (see **Fig. 4a**). However, if the inference of $DNN_{Quantification}$ predicted < 40 pg/mL, contradicting the prediction of $DNN_{Classification}$ for the same sample, we then used a second quantification model, namely $DNN_{Low}$, which was treated as an adjudicative model. $DNN_{Low}$ was only used if the classification and quantification neural networks disagreed with each other and the blind inference of $DNN_{Low}$ was used as our final concentration measurement if it was ≥ 40 pg/mL - i.e., complying with the inference of $DNN_{Classification}$ (see Supplementary Note 4). If the prediction from $DNN_{Low}$ also disagreed with the inference of $DNN_{Classification}$, the sample was then labeled as "undetermined" and excluded from the quantification results. Therefore, the final cTnI quantification was implemented with this collaboration among the three neural networks, as illustrated in **Fig. 4a**.

$DNN_{Low}$ was trained separately from $DNN_{Quantification}$ using samples with lower cTnI concentrations (i.e., < 1,000 pg/mL), whereas $DNN_{Quantification}$ was trained across a larger concentration range of 40–40,000 pg/mL. Quantitative cTnI predictions by the optimized neural network models for the samples from the *validation* set showed a strong correlation with the ground truth concentrations measured by the benchtop device, achieving a Pearson's *r* of 0.962 (see **Fig. S9b**, and the "Neural network-based analysis" subsection of the Methods for details of the neural network training and architectures).

After optimizing the cTnI inference models using validation serum samples, the resulting optimal neural network models were blindly tested on a set of 46 serum samples from 23 patients - never seen before. Serum samples from each patient were measured in duplicate with two separate hs-VFAs activated per patient (**Table S1**). Blind testing results for the classification model (with a clinical cut-off of 40 pg/mL) showed a sensitivity and a specificity of 97.1% and 83.3%, respectively, with 1 false negative (FN) and 2 false positive (FP) predictions (**Fig. 4b**). These FN and FP predictions can be attributed to the potential degradation of the clinical samples over storage time, matrix effects, low-cost design of the VFA cartridge or the small size of the model training set[47]. We should emphasize that the performance of the classification model could be further improved by duplicate sample tests – at the cost of an increase in the sample volume per test. For instance, the sample can be labeled as "undetermined" if the variation in the predicted concentrations between the duplicate tests exceeds a certain threshold. This threshold was empirically determined by analyzing the mean absolute error (MAE, see the "Statistical analysis" subsection of the Methods) between the duplicate tests of each sample. We found out that the MAE scores for all 3 falsely classified samples (i.e., 1 FN and 2 FPs) were larger (i.e., > 10%) than for the correctly labeled samples (**Table S2**). Therefore, with a test-to-test variation threshold of ~10%, we could reduce the number of FNs and FPs to 0 and improve the accuracy of $DNN_{Classification}$ to 100% by correcting the falsely classified samples as "undetermined". Due to the requirement of duplicate testing, we did *not* employ this strategy in the rest of our clinical testing analysis and kept the serum sample volume at 50 µL per test.

After this initial blinded cTnI classification stage, 35 serum samples that were classified into ≥ 40 pg/mL range, including 33 correctly classified samples (from 17 patients) and 2 falsely classified samples (from 1 patient), were further processed by $DNN_{Quantification}$. Out of these 35 samples, 21 were only quantified by $DNN_{Quantification}$ revealing the final concentration measurements; the remaining 14 had contradicting predictions between $DNN_{Quantification}$ and $DNN_{Classification}$, and therefore were also processed by $DNN_{Low}$ - the adjudicative quantification model. Since none of the 14 predictions from $DNN_{Low}$ disagreed with $DNN_{Classification}$ (i.e., all concentration predictions from $DNN_{Low}$ were in the ≥ 40 pg/mL range) no serum samples from the blind testing set were labeled as "undetermined". The final quantitative predictions of cTnI concentrations from the two quantification neural network models on these 35 samples had a good match with the gold standard measurements from an FDA-cleared analyzer, achieving Pearson's *r* of 0.965. In addition, the mean variation between duplicate measurements of the samples was minimal, with an average CV of 6.2%, confirming a repeatable inter-sensor operation (**Fig. 4c** and **Table S1**).

Importantly, cTnI quantification by the neural network-based method outperformed a standard rule-based method where cTnI concentration and the time-lapse hs-VFA signal are related to each other with an explicit equation. The optimal equation from this rule-based method for our case represented a power law fit (see Supplementary Note 5). To provide a fair comparison, the deep learning models and the power law function were created using the same training set, and then tested on identical blinded serum samples. For the same test samples, the optimized quantification network models ($DNN_{Quantification}$ and $DNN_{Low}$) outperformed the power-fitting model in terms of accuracy and reproducibility (i.e., Pearson's *r*: 0.861 vs. 0.965 and CV: 10.5% vs. 6.2%; **Fig. S10a**). The advanced performance of neural network models originates from their inherent universal function approximation power and ability to effectively learn robust quantification functions between the analyte concentration and non-linear time-lapse response of hs-VFA despite the interference of noise from clinical samples and the low-cost design of the paper-based POC sensor.

In addition, the performance of the cTnI quantification by neural network models improved from the incorporation of the OA step and the use of the time-lapse imaging method. For instance, cTnI quantification precision achieved by the optimized deep learning models ($DNN_{Quantification}$ and $DNN_{Low}$) using the time-lapse signal inputs processed by OA (i.e., $I_{Time-lapse}$) was higher compared to the precision achieved by the same models using the time-lapse inputs *without* OA (see **Fig. S10b**, a CV of 8.8% for time-lapse inputs without OA vs. a CV of 6.2% for time-lapse inputs with OA). In addition, the cTnI concentrations predicted by optimized models with the end-point inputs (i.e., at t = 60 s) had a larger deviation from the ground truth values than the predictions from the same models with the time-lapse inputs (**Fig. S10c**). This was especially apparent in the lower concentration range. In the models with end-point and time-lapse inputs, Pearson's *r* coefficients were 0.806 and 0.959, respectively, for concentrations below 1,000 pg/mL (**Table S3**). Therefore, both the time-lapse imaging and computational assay quality check (i.e., OA) have a positive impact on the quantification performance of the neural network models, and the incorporation of these methods into the neural network-based analysis is important for more accurate and robust quantification of cTnI concentrations.

**Discussion**

cTnI concentrations quantified by our hs-VFA platform using neural network-based analysis showed a strong correlation with the ground truth measurements obtained by an FDA-cleared analyzer (achieving a Pearson's *r* of 0.965) and demonstrated competitive testing precision with an average CV of 6.2%, which falls within the precision requirement of the clinical hs-cTnI assays (i.e., CV of <10%)[12]. In addition, the neural network-based algorithms were able to quantify cTnI levels over a large concentration range and correctly classify samples above/below the clinical cut-off (40 pg/mL). This successful performance of hs-VFA can be attributed to the synergistic effects of several key innovations: (i) a signal amplification reaction which provided a 212.3-fold increase in the sensitivity threshold of the assay; (ii) time-lapse imaging which provided an additional 6.5-fold increase in the sensitivity threshold; (iii) computational assay quality check algorithms, including VCI and OA, which provided a 2.1-fold improvement in LoD also reducing the CV of the assay; and (iv) neural network-based cTnI inference approach which further helped reduce the CV. Moreover, the compact and cost-effective paper-based sensor ($3.86 per test, **Table S4**), paired with a custom-designed Raspberry Pi-based user-friendly portable reader (priced at approximately $170 per prototype, **Table S4**), emphasizes the suitability of this platform for POC assays and positions it as a viable alternative for standard laboratory testing. When benchmarked against commercial cTnI-POCT assays and recent research results in the literature (**Table 1**), our hs-VFA has a competitive precision over a wide assay range, spanning 6 orders of magnitude in concentration, fully covering the clinically relevant cTnI range (0.01 – 100 ng/mL)[18]. It also provides better sensitivity, with an order of magnitude lower LoD. The improved performance of hs-VFA could be leveraged to enable 0-hour rule-out with a single cTnI test in low-risk patients[48] and improve the prehospital phase of care for high-risk patients empowered by accurate cTnI testing even in primary care or nursing home infrastructure[6].

The quantification of the analyte concentration in clinical samples conventionally relies on calibration curves established using, e.g., spiked samples with known antigen concentrations. However, this method has inherent limitations that may compromise the accuracy and adaptability of the assay. For example, when applying this calibration-based approach to the hs-VFA clinical sample results, we observed a significant performance degradation, as shown in **Fig. S10a**. This limited performance in calibration-curve-based cTnI quantification may arise due to various factors, including (i) matrix effects from the spiked and clinical samples, (ii) varying storage conditions for different sample batches, (iii) variations in the assembly between sensor batches, and (iv) limited control over environmental factors (i.e., temperature, humidity). Furthermore, ensuring the accuracy of the calibration curve over time can be challenging, necessitating frequent recalibration and quality control measures. In contrast, neural network-based concentration inference can adapt to more diverse sample matrices and variations in sample/sensor batches better than a fixed function determined with a single explicit rule[39].

We should also note that the introduction of the Raspberry Pi system for constructing the portable reader yields multiple benefits compared to using a smartphone-based reader. Our custom-designed Raspberry Pi-based reader showed equivalent quantification performance to a smartphone-based reader[39-42] (**Fig. S11**). The Raspberry Pi's open-source ecosystem delivered higher flexibility and customization options in hardware and software, facilitating the operation of the device for time-lapse imaging-based analysis. The standalone nature of the Raspberry Pi-based reader can help unify the design to a fixed footprint and camera model, contrary to smartphone-based readers, which are sensitive to the model brand and camera optics, which frequently change as new models are introduced in the market. Furthermore, our Raspberry Pi reader offers cost benefits compared to high-end smartphones. The Raspberry Pi-based custom-designed hs-VFA reader can be easily equipped with batteries and communication modules (3G/4G/5G/Wi-Fi), fulfilling the real-time connectivity requirements of POCT[49]. This could help transform the hs-VFA platform into a more advanced POCT assay that is not only used in traditional medical facilities but also enables cTnI measurements to be performed during patient transport (e.g., ambulance-based testing)[50], thus expanding the potential diagnostic use cases.

One of the major advances behind the presented hs-VFA results is the highly sensitive detection of a single biomarker with an LoD of < 1 pg/mL. To increase the analytical sensitivity, we designed the sensing membrane to improve the overall surface area for capturing the target analyte, coupled with a signal amplification reaction. With this strategy, we configured the membrane to conduct 10 repeated tests under equivalent conditions across 17 multiple spots in a single assay run (all within 15 min), which is not achievable with conventional rapid diagnostic tests such as lateral flow assays. When investigating the impact of the number of test spots on the assay sensitivity without data refinement, we observed that increasing the number of test spots enhanced the signal intensity and assay sensitivity (**Fig. S12a–b**). Nevertheless, the assay uniformity (**Fig. S12c**) and repeatability (**Fig. S12d**) were reduced due to spot-to-spot variability between the inner (closer to the center) and outer (closer to the edges) reaction spots on the sensing membrane. We resolved these spot-to-spot and test-to-test variation issues through VCI and OA by taking advantage of the statistical benefits derived from the redundancy of multiple test spots on the test membrane, similar in concept to "swarm sensing"[51]. Our experiments on cTnI spiked/clinical sample assays showed that excluding the compromised spots ensures data integrity, achieving more reliable and consistent assay results without a trade-off in assay sensitivity (**Fig. 3b–d**). This approach mitigates errors inherently associated with paper-based sensors (due to the use of inexpensive components such as paper membranes, the possibility of nonuniformity in fluid dispersion, and potential defects during the antibody spotting and sensor assembly processes), demonstrating that high-sensitivity analyte measurements at the pg/mL level can be reliably achieved using the low-cost assay platform of our hs-VFA. Additionally, incorporating OA during the assay development and validation stages has the potential to reduce sensor production costs by leveraging the spot position-dependent exclusion rate (**Fig. 3, insets i–ii**). For example, by selecting 6 test spot positions that consistently exhibit a < 30% exclusion rate, we anticipate a 36% reduction in the cost of antibodies (equivalent to $0.7 per test). This reduction is significant, as antibodies purchased at low volumes account for > 50% of the unit sensor cost (**Table S4**); we should also emphasize that over large-volume manufacturing practices, the overall cost per test can be reduced to < $1 per test.

In summary, we developed a deep learning-enhanced paper-based hs-VFA platform that achieved highly sensitive (LoD, 0.2 pg/mL) and precise (average CV < 7%) quantification of cTnI in serum samples within 15 min per test, showcasing significant progress in advancing high-performance POC sensors. Notably, the performance enhancement of hs-VFA platform was achieved without compromising key attributes inherent to POCT, such as low cost, simple operation, rapid assay times, and digital connectivity, which underscore the significance of our study. Our results highlight the potential to democratize diagnostic testing by demonstrating that high-quality assays, traditionally limited to high-end clinical analyzers in central laboratories, can be performed on cost-effective POC platforms. We envision this technology being widely used to expedite global diagnostic equity for testing challenging disease biomarkers.

## Methods

**Antibody conjugation to AuNP**: Antibody conjugation to 15 nm AuNP is based on physical adsorption. In brief, the anti-cTnI detection antibody (10 μL, 1 mg/mL; 19C7, Hytest) was suspended in 15 nm AuNPs (1 mL; BBI Solutions) mixed with 100 mM borate buffer (100 μL, pH 8.5; Thermo Scientific). Following a 1 h incubation using a rotary mixer (20 rpm) at room temperature (RT), the conjugate was blocked for 2 h by adding bovine serum albumin (BSA; 10 μL, 10% w/w; Thermo Scientific). Then, the conjugate underwent three rounds of centrifugal washing (21380g, 25 min, 4 °C) using a 10 mM borate buffer (pH 8.5). Next, the final conjugate pellet was resuspended in the storage buffer (100 μL) containing 5% w/w trehalose (Sigma), 0.5% w/w protein saver (Toyobo), 0.2% v/v Tween 20 (Sigma), and 1% v/v Triton X-100 (Sigma) in 10 mM PBS (pH 7.2; Thermo Scientific). The absorption spectra and concentration of the conjugate were analyzed using a microplate reader (Synergy H1; BioTek). The conjugates were stored at 4 °C at 9 OD concentrations until ready for use.

**Sensing membrane preparation and hs-VFA assembly**: The sensing membrane fabrication involves four steps: printing, heating, antibody spotting, and blocking. A wax printer (Xerox) was used to print and compart 17 reaction areas outlined by a black background onto a nitrocellulose membrane (NC, 0.2 μm; Bio-Rad). The wax-printed membranes were baked (120 °C, 55s) in a forced air convection oven (Across International) to melt the wax on the membrane's top side. This created a three-dimensional compartment with hydrophilic reaction areas against a hydrophobic background. Considering heat convection inside the baking chamber, we uniformly prepared up to 30 sensing membranes (arranged in a 6 × 5 array, with 1 mm gaps between membranes) in a single baking batch. The anti-cTnI capture antibody (0.8 μL, 1 mg/mL; 560 Hytest) and goat anti-mouse IgG (0.8 μL, 0.1 mg/mL; Southern Biotech) were respectively dispensed onto the test and positive control spots. The batch membrane sheet was subsequently dried in the oven (37 °C, 15 min). The membrane sheet was then immersed in 1% w/w BSA solution for blocking (30 min, at RT). After another round of drying the membrane sheet (37 °C, 15 min), it was divided into individual sensing membranes using a razor.

All the paper materials for hs-VFA were prepared as previously outlined[40]. Briefly, raw paper materials were precisely cut using a CO$_2$ laser cutting system (Trotec). The top case for immunoassay (1$^{st}$ top) was assembled by sequentially stacking engineered paper layers (absorption layer, flow diffuser, 1$^{st}$ spreading layer, interpad, 2$^{nd}$ spreading layer, and supporting layer) with a double-sided adhesive foam tape as an assembly frame. A concentric circular pattern in the flow diffuser and outer contour in the supporting layer were prepared using the same wax-printing/baking process as the sensing membrane. The flow diffuser, interpad, and supporting layer were treated with 1% w/w BSA solution for blocking. The bottom case was prepared by stacking five absorption pads and affixing the sensing membrane on top using adhesive foam tape.

**hs-VFA cartridges**: The plastic cartridges for hs-VFA were 3D-printed using Form 3 (gray resin, Formlabs) with 100 μm resolution mode. The design of the bottom and 1$^{st}$ top cases aimed to compress the stacked paper materials by 25% of the initial total paper layers' thickness, enhancing flow diffusion, assay uniformity, and efficiency (**Fig. S4a**). For the 2$^{nd}$ top case used for signal amplification, a transparent acrylic window (16.3 mm × 16.3 mm × 1 mm, laser-cut) was affixed to the 3D-printed cartridge using a clear acrylic adhesive. Foam tape was employed as a gasket to prevent reagent leakage in the 2$^{nd}$ top case. The transparent window has four ventilation outlets (0.8 mm in diameter) at each corner. These outlets are designed to remove air from the reaction chamber, facilitating the inflow of reagent solution. The reagent inlet of the 2$^{nd}$ top case was designed to hold a volume of up to 1 mL of solution.

**Portable reader and image analysis**: A custom-designed portable reader was assembled using 3D-printed parts (i.e., housing, LED holder, and cassette tray) and low-cost off-the-shelf components (i.e., LEDs [DigiKey], macro lens [Edmund Optics], Raspberry Pi, camera module V2-8 megapixel [Adafruit], and touch screen display [Elecrow]). The 3D-printed custom parts were produced by Object 30 (Stratasys) and Ultimaker 3 (Ultimaker) 3D printers. The LED module contained four green LEDs (532 nm) for time-lapse imaging and two white LEDs for brightfield imaging arranged in a circular shape. All LEDs were polished from the front to provide even light distribution across the sensing membrane. The reader was designed for easy pedestal installation with four optical posts, enabling both hand-held and benchtop use. The user interface of the image capture software consisted of a real-time camera screen and input fields with user-adjustable parameters (i.e., exposure time, the number of images to capture, and capture interval, **Fig. S1**) for automated time-lapse imaging. All images were captured in raw format under consistent imaging conditions (100 μs exposure time, and 30 s interval).

After the end of the hs-VFA operation (per test), captured time-lapse images (*N*=3) were processed by an automated image segmentation code that extracted the green channel from the RGB images, segmented all 17 immunoreaction spots from the activated sensing membrane and averaged pixels within each segmented spot to generate 17 intensity values per image ($s_i^{j,t}$, i ∈ {Test, Pos, Neg}– type of immunoreaction, j– spot repeat within the given type, t– time point). These intensity values of time-lapse images were further normalized by the corresponding background intensities ($b_i^{j,t}$) from the sensing membrane image captured before the start of the hs-VFA operation and raw absorption signals for each time-lapse image were calculated as:

$$I_i^{j,t} = 1 - \frac{s_i^{j,t}}{b_i^{j,t}}. \tag{5}$$

Alike spots within each immunoreaction type were averaged, resulting in a total of 3×N raw absorption signals ($I_i^t$) per assay, where $N=3$ is the total number of captured images during time-lapse operation.

**Assay operation**: Assay operation comprises two main steps: immunoassay and signal amplification. For the immunoassay, the bottom case is assembled with the 1st top case. The process begins by activating the device with the addition of 1st running buffer (200 µL), which consists of 1% v/v Triton X-100 and 1% v/v BSA in PBS (10 mM, pH 7.2). After allowing 30 s for complete absorption to paper layers, a mixture (100 µL) of the sample (serum, 50 µL) and conjugate (2.5 OD, 50 µL) is added following immediate mixing. This mixture is left for 1 min to ensure complete absorption. Next, 2nd running buffer (300 µL) is introduced. This buffer contains 3% v/v Tween 20, 1% v/v albumin, 0.5% w/w protein saver, and 1% w/w trehalose in PBS (10 mM, pH 7.2). Its purpose is to maintain the flow of solutions within the hs-VFA structure and to wash away any unbound conjugate and target molecules from the sensing membrane, thereby minimizing non-specific binding. The 1st top case is removed after 8.5 minutes and replaced with the 2nd top case for Au-ion reduction-based signal amplification. In this step, 500 µL of a reagent solution containing 10 mM HAuCl$_4$ (Sigma) and 10 mM H$_3$NO (Sigma) in PBS is added to supply the reagent to the sensing membrane and absorption pads. After 3 min, when the reagent feeding is complete, the 2nd top case is removed, and the bottom case is transferred to the reader for time-lapse imaging.

**cTnI spiked and clinical serum samples**: The cTnI spiked serum samples were prepared for assay optimization and validation by spiking cTnI standard antigen (I-T-C complex purified from the human heart; Lee Biosolutions) into cTnI-free serum (Hytest). To obtain various concentrations of cTnI, we performed serial dilutions using the same serum. The cTnI antigen was stored at -80°C after being aliquoted into 1 µL portions. A fresh solution was prepared immediately before each assay to prevent potential antigen degradation. Clinical serum samples containing cTnI were provided by UCLA Health. This study was approved by the UCLA Institutional Review Board (IRB no. 20-002084). Patient consent was waived since these specimens were pre-existing remnant samples collected independent of this research project. The ground truth values of the clinical samples were determined using a standard analyzer (Access, Beckman Coulter) at UCLA Health, immediately after collection. The analyzer had a cut-off level of 40 pg/mL, which quantified samples with cTnI levels ≥ 40 pg/mL and displayed results below the cut-off as < 40 pg/mL. We tested 54 clinical samples, including 35 samples with cTnI levels ≥ 40 pg/mL and 19 samples with cTnI levels < 40 pg/mL (**Table S1**). Additionally, we tested 8 serially diluted clinical samples derived from two stocks. Clinical samples were stored at -80°C and were thawed immediately before the assay at 4°C for 1 h.

**Computational analysis of hs-VFA signals**: Prior to neural network-based analysis, all the assays activated during clinical sample testing underwent computational assay validation, which consisted of two steps: VCI and OA. At the VCI step, individual assays were excluded from the clinical sample dataset based on the ratio between raw intensities from positive ($I_{Pos}^t$) and negative ($I_{Neg}^t$) control spots. An assay was excluded if the ratio did not fall into the CI calculated according to:

$$r_{Low}^t < \frac{I_{Pos}^t}{I_{Neg}^t} < r_{High}^t, \; t = 1...N, \tag{6}$$

where $r_{Low}^t$ and $r_{High}^t$ are calculated as ±1.96 SD from the mean of the distribution of the positive and negative control ratios of 120 assays (i.e., 80 spiked samples + 40 clinical samples), $N=3$ is the total number of captured images during time-lapse operation.

Next, during the OA step, individual test spots from each assay were excluded based on 95% acceptance range from the statistical distribution of the $N_{Test} = 10$ test spots:

$$\frac{<I_{Test}^{j,t}>-2.262*\sigma(I_{Test}^{j,t})}{\sqrt{N_{Test}}} < I_{Test}^{j,t} < \frac{<I_{Test}^{j,t}>+2.262*\sigma(I_{Test}^{j,t})}{\sqrt{N_{Test}}}, \qquad (7)$$

j=1…$N_{Test}$, t = 1…N – time point, 2.262 is the t-score for a two-tailed test with 9 degrees of freedom (i.e., $N_{Test} - 1$). Averages ($<I_{Test}^{j,t}>$) and SDs ($\sigma(I_{Test}^{j,t})$) were calculated over the test spot repeats on the same paper-based test membrane (i.e., j=1…10) independently for each of the 3 membrane images captured during the time-lapse operation.

In addition to OA based on absolute test spot values (termed statistical OA model), we also performed OA based on differential spot values (i.e., OA-D model). The 95% acceptance range for OA-D was defined as:

$$\frac{<I_{Test,Diff}^{j,t}>-2.262*\sigma(I_{Test,Diff}^{j,t})}{\sqrt{N_{Test}}} < I_{Test,Diff}^{j,t} < \frac{<I_{Test,Diff}^{j,t}>+2.262*\sigma(I_{Test,Diff}^{j,t})}{\sqrt{N_{Test}}}, \qquad (8)$$

$I_{Test,Diff}^{j,t} = I_{Test}^{j,t+1} - I_{Test}^{j,t}$, j=1…$N_{Test}$, t = 1…N-1. Uniformity and reproducibility of hs-VFA after applying the statistical OA model showed superior improvement compared to CV before any OA and CV after applying the OA-D model (**Fig. 3b**, Supplementary Note 3); therefore, we selected the statistical OA method for data refinement of the clinical samples. Refined hs-VFA signals after OA are defined as $I_{i,Ref}^t$.

Time-lapse response from hs-VFA using the OA-refined signals ($I_{i,Ref}^t$) was further calculated according to equation (1) (i.e., $I_{Time-lapse}$). During the sensor optimization and testing on spiked samples (**Fig. 3c–d**), time-lapse signals ($I_{Time-lapse}$) were normalized by the time-lapse signal from the negative control sample generating normalized time-lapse signals ($I_{Normalized,\ Time-lapse}$, see equation (2)). For the clinical sample test (**Fig. 3e**), we used time-lapse signals without normalization ($I_{w/o\ Normalized,\ Time-lapse} = I_{Time-lapse}$) in order to minimize variations between different testing batches.

**Neural network-based analysis:** The neural network-based analysis consisted of classification and quantification parts and a total of three independent shallow fully-connected neural networks, including one network for classification between samples from ≥ 40 pg/mL and < 40 pg/mL concentration ranges (i.e., $DNN_{Classification}$), and two networks for quantification of cTnI in ≥ 40 pg/mL range (i.e. $DNN_{Quantification}$ and $DNN_{Low}$) - see **Fig. 4a**. Inputs to all three neural networks represented the time-lapse hs-VFA signals calculated from the three time-lapse images according to equation (1) (i.e., $I_{Time-lapse}$). The input signal was standardized according to the formula:

$$I_{Time-lapse}^{st} = \frac{I_{Time-lapse}-<I_{Time-lapse}>}{\sigma(I_{Time-lapse})}, \qquad (9)$$

where $<I_{Time-lapse}>$ and $\sigma(I_{Time-lapse})$ are the mean and the SD of the time-lapse signal, respectively, calculated over the training dataset.

The classification network ($DNN_{Classification}$) consisted of 3 hidden layers (512, 256, 128 units) with 'ReLU' activation functions and L2 regularization for all layers (see **Fig. S7**). Each hidden layer was followed by a batch standardization layer. The output layer had 1 unit with a 'sigmoid' activation function. The loss function that we used was binary cross-entropy compiled with Adam optimizer, a learning rate of 1e-4 and a batch size of 5. Binary cross-entropy loss ($L_{BCE}$) is defined as:

$$L_{BCE} = -\frac{1}{N_b}\sum_{i=1}^{N_b}(y_i \log(y'_i) - (1-y_i)\log(1-y'_i)), \qquad (10)$$

where $y_i$ are the ground truth labels (i.e., "1" for samples from $\geq$ 40 pg/mL concentration range and "0" for samples from < 40 pg/mL concentration range), $y'_i$ are the predicted labels and $N_b$ is the batch size.

The quantification part consisted of two independent fully-connected neural networks (namely $DNN_{Quantification}$ and $DNN_{Low}$). $DNN_{Quantification}$ contained 3 hidden layers (256, 128 and 64 units), each with 'ReLU' activation functions and L2 regularization. $DNN_{Low}$ consisted of 2 hidden layers (128 and 64 units), each with 'ReLU' activation functions and no regularization (**Fig. S8**). Each hidden layer in both models was followed by a batch standardization layer. The loss function for both models was the mean squared logarithmic error (MSLE) compiled with Adam optimizer, a learning rate of 1e-4 and a batch size of $N_b$ =5. MSLE loss is defined as:

$$MSLE = \frac{1}{N_b}\sum_{i=1}^{N_b}(log(y_i+1) - log(y'_i+1))^2, \qquad (11)$$

where $y_i$ are the ground truth cTnI concentrations, $y'_i$ are the predicted concentrations. Both of these quantification models ($DNN_{Quantification}$ and $DNN_{Low}$) predicted cTnI concentrations ($y'_i$) in pg/mL. Architectures of all three neural networks were optimized through a 4-fold cross-validation on the validation sets.

The classification model ($DNN_{Classification}$) was trained on 64 serum samples from 32 patients and validated on 42 samples from 21 patients (i.e., serum from each patient was tested in duplicates with two activated hs-VFAs per patient). Out of 64 training samples 54 were clinical samples and the remaining 10 were diluted samples. The optimal classification network ($DNN_{Classification}$) achieved 84.6% sensitivity and 93.8% specificity on the validation set (**Fig. S9a**). The optimized model was further blindly tested on 46 additional serum samples from 23 patients, achieving a sensitivity and specificity of 97.1% and 83.3%, respectively, as reported in the Results section.

The quantification stage was only used when $DNN_{Classification}$ blindly classified the sample as $\geq$ 40 pg/mL. $DNN_{Quantification}$ was trained on 64 samples from 32 patients and validated on 22 samples from 11 patients. cTnI concentrations in the training samples were in 40–40,000 pg/mL range, while cTnI concentration range for the validation set was 50–5,000 pg/mL. Quantification results from $DNN_{Quantification}$ revealed the final cTnI concentration measurement per sample if and only if its blind inference agreed with $DNN_{Classification}$ inference (see **Fig. 4a**). For 10 out of 22 validation samples, the blind inference results of $DNN_{Quantification}$ were in < 40 pg/mL range, contradicting the predictions from the classification model (i.e., all of these samples were classified into $\geq$ 40 pg/mL range by $DNN_{Classification}$). As illustrated in **Fig. 4**, such samples were further processed by a second quantification network, $DNN_{Low}$, which was used as an adjudicative model. If the blind inference from $DNN_{Low}$ still disagreed with $DNN_{Classification}$, the sample was labeled as undetermined; otherwise the prediction from $DNN_{Low}$ was used as the final concentration measurement result for the cases where $DNN_{Classification}$ and $DNN_{Quantification}$ disagreed with each other. $DNN_{Low}$ was trained separately from $DNN_{Quantification}$ on 34 serum samples (from 17 patients) with cTnI ground truth levels below 1,000 pg/mL and validated on 10 samples (from 5 patients) with cTnI concentrations in 50–200 pg/mL range.

One out of 10 validation samples processed by $DNN_{Low}$ had < 40 pg/mL concentration prediction and it was labeled as "undetermined". Predictions of the optimized quantification models on the rest of the validation samples (i.e., 21 samples) had a high correlation with the ground truth cTnI concentrations quantified by an FDA-cleared analyzer with a Pearson's *r* of 0.962 (**Fig. S9b**). For the blind testing set composed of 18 new serum samples, these networks achieved a Pearson's *r* of 0.965 as reported in the Results section. Training sets for both $DNN_{Quantification}$ and $DNN_{Low}$ included samples in < 40 pg/mL range and during the training process, MSLE loss for such samples was increasing only if the predicted concentration was ≥ 40 pg/mL. Incorporation of the samples under 40 pg/mL cut-off into the training sets helped to create more robust quantification models and achieve reliable cTnI quantification despite a limited number of training samples.

**Statistical analysis**: For assay validation and cTnI spiked serum sample tests, all experimental data were presented as the mean of at least three measurements ± SD. Clinical sample test results were derived from the mean of duplicate measurements ± SD. Further details regarding experimental replicates are provided in the corresponding figure legends. The CV was calculated by dividing the SD by the mean (%). Group differences were assessed using an unpaired two-sample *t*-test, with statistical significance set at *P* < 0.05. In addition to CV, we used MAE to estimate repeatability between duplicated samples, defined as:

$$MAE = \frac{|Rep_1 - Rep_2|}{Rep_1 + Rep_2}, \tag{12}$$

where $Rep_1$ and $Rep_2$ are first and second repeats of the serum sample respectively.

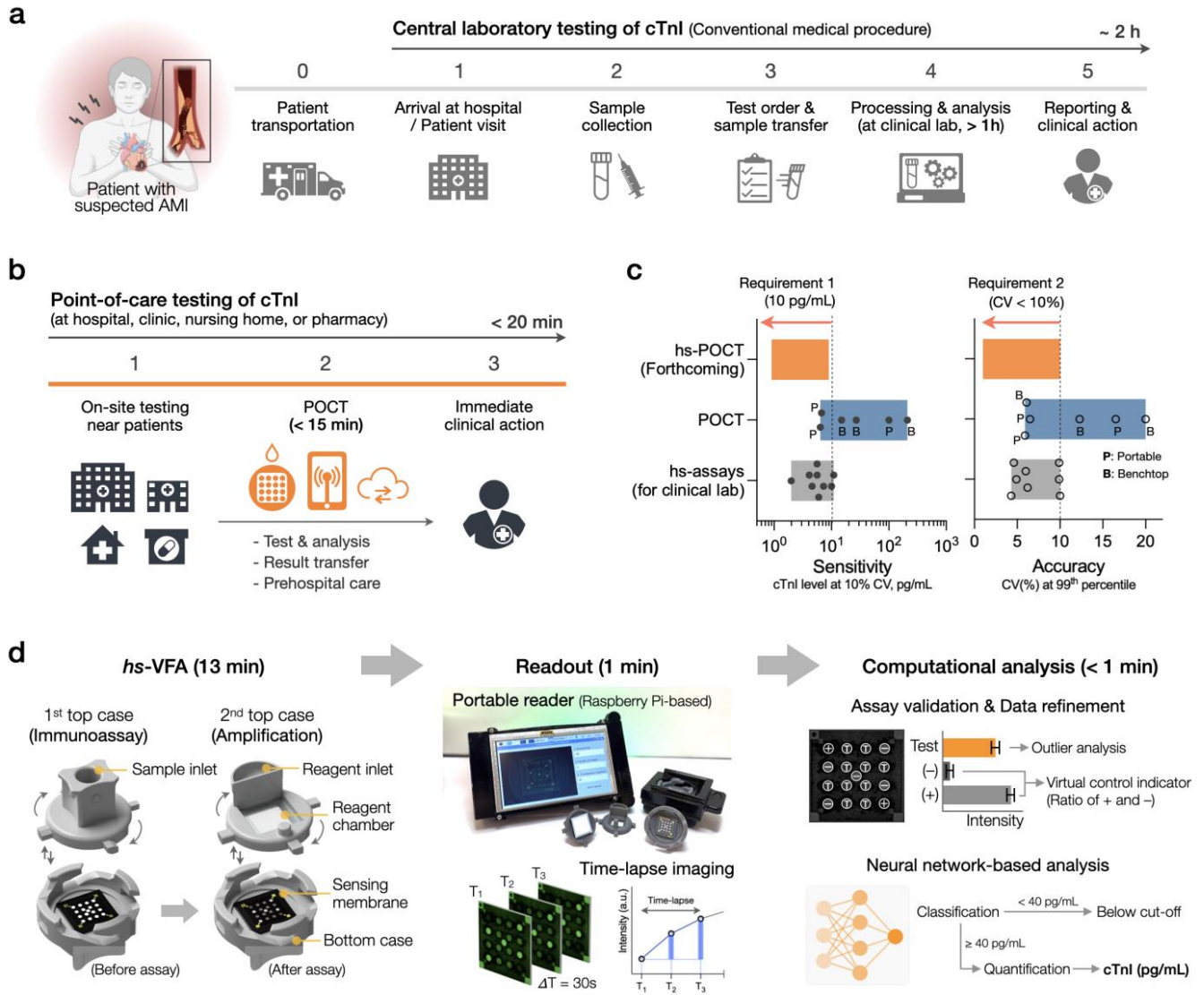

**Fig. 1.** Overview of paper-based hs-VFA and its application in hs-POCT of cTnI. (a) Traditional workflow of central laboratory tests for diagnosing patients with AMI. (b) POCT using hs-VFA and its mobile reader system can enhance the efficiency and quality of patient care. (c) Current performance landscape of commercial cTnI assays (hs-assays and POCT assays) and the requirements for a successful hs-POCT cTnI assay. The bars show the ranges of sensitivity and precision specifications, indicated by dots and circles, respectively. (d) Schematic summary of our approach for hs-POCT cTnI assay using hs-VFA, a portable reader, and computational analysis.

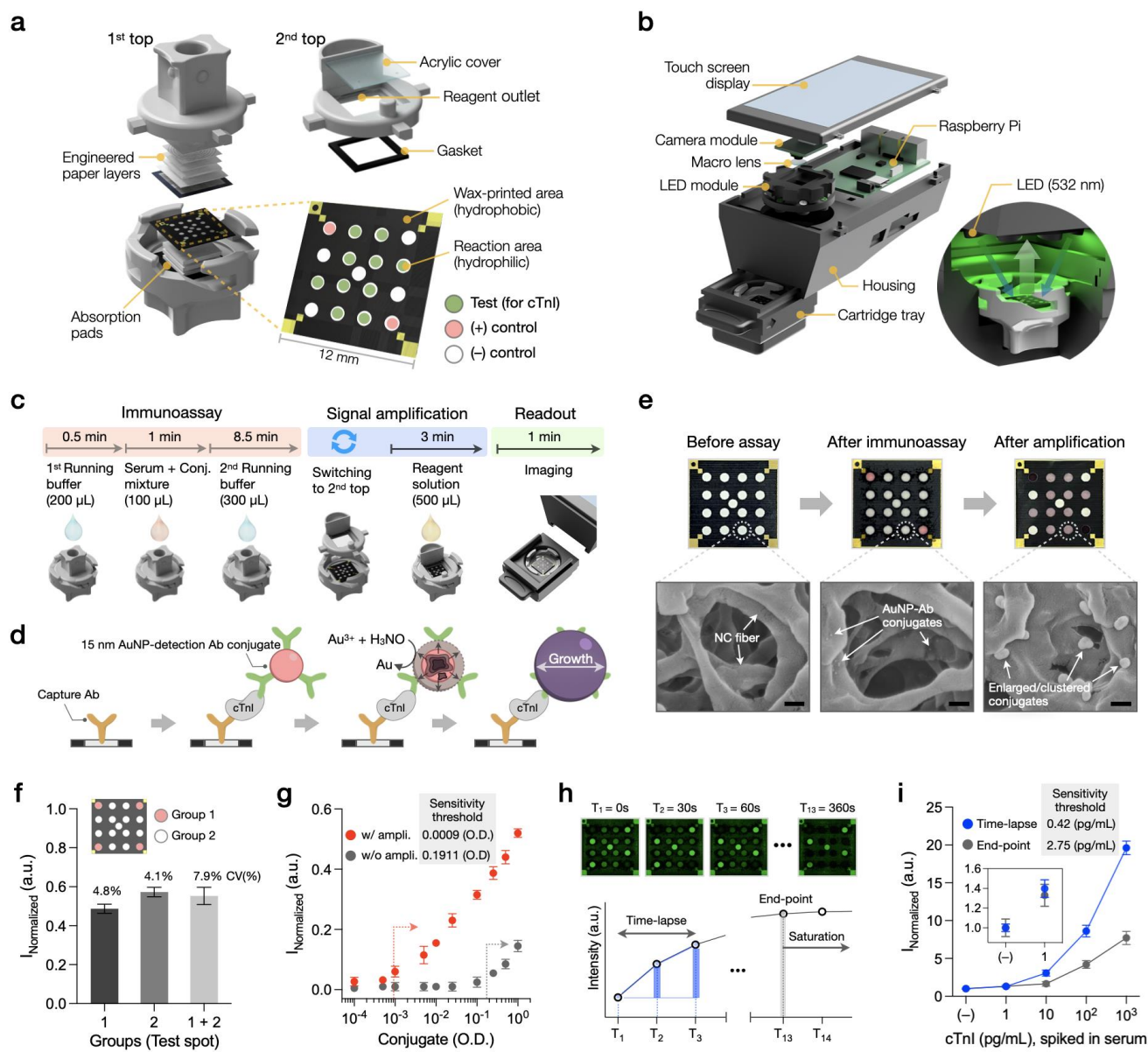

**Fig. 2.** Assay, reader, and imaging strategy of the hs-VFA. (a) Structure of the hs-VFA, comprising two top cases and a bottom case with a sensing membrane. (b) Portable reader based on Raspberry Pi. The inset displays the internal cassette tray during imaging under LED illumination. (c) Assay process and timeline. (d) Principle of immunoassay and signal amplification based on the reduction of $Au^{3+}$ onto the surface of AuNP conjugates. (e) Sensing membrane images at each assay step and corresponding SEM images of the test spots. Scale bar: 200 nm. (f) Evaluation of spot uniformity in the amplification reaction. Three individual sensors treated with 1 OD AuNP-antibody conjugate in each spot were used for comparison. (g) Effect of the signal amplification on analytical sensitivity. Conjugates were serially diluted and pre-immobilized on the sensing membrane. Two top-case operations were conducted. (h) Comparison of imaging methods: time-lapse vs. end-point analyses. (i) Impact of time-lapse imaging on assay sensitivity. Data points in (g) and (i) represent the mean of triplicates ± SD.

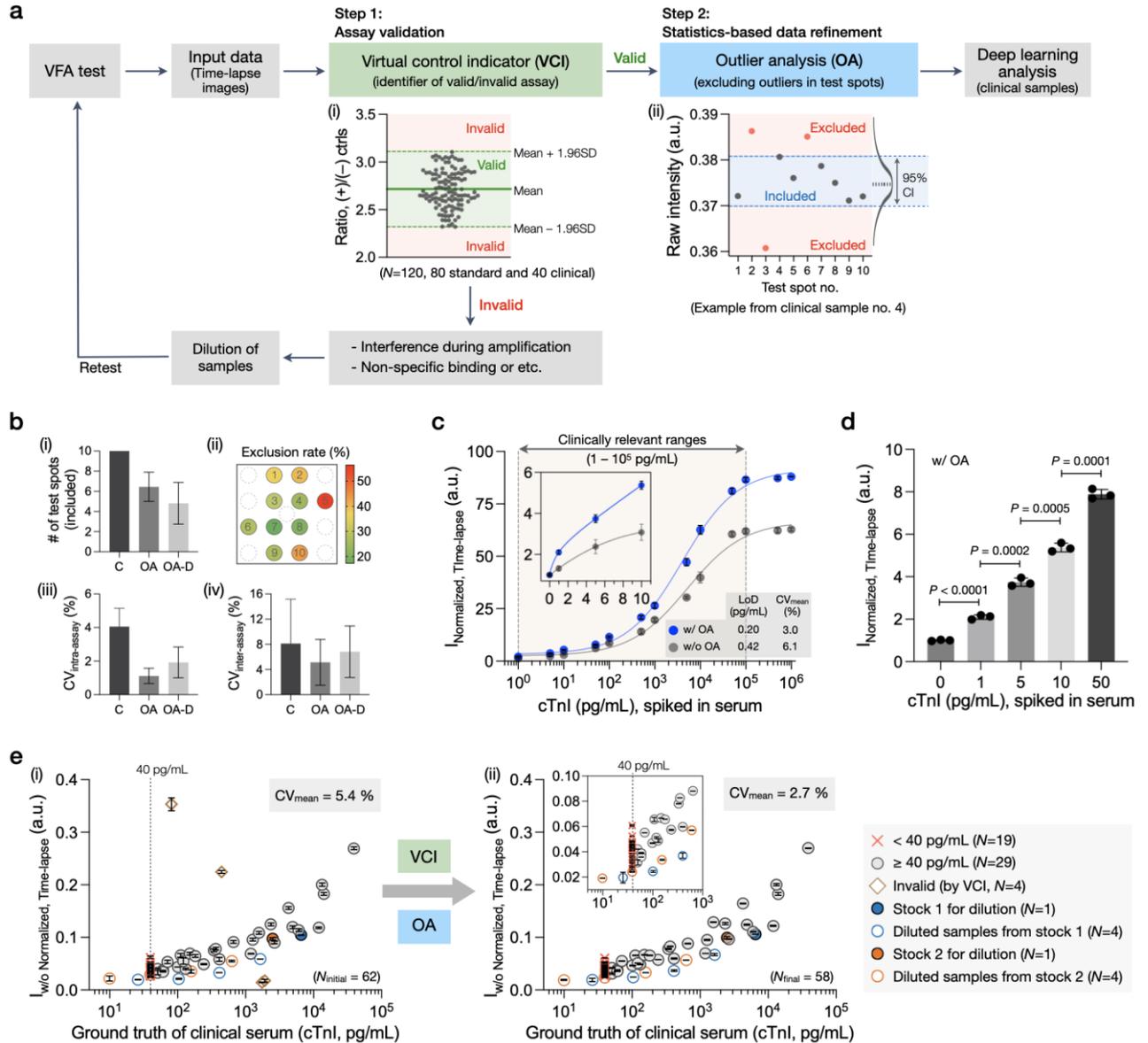

**Fig. 3.** Computational data processing and the impact of assay quality check algorithms for cTnI testing using hs-VFA. (a) Data-processing pipeline; (i) Dataset and criteria to create a VCI. (ii) An example illustrating data refinement through OA. (b) Effect of OA; (i) The number of remaining test spots. (ii) Exclusion rate for each test spot. (iii) Test spot uniformity within a single cassette (intra-assay). (iv) Assay repeatability (inter-assay). Data in (i)–(iii) are based on $N$=330 (55 spiked and 55 clinical samples with three time points). Data in (iv) are based on $N$=21 (7 cTnI spiked serum samples, triplicates). C stands for control, before OA. OA refers to a statistics-based method and OA-D refers to a differential-based method (see the Methods section). (c) Test results for cTnI spiked serum sample and calibration plot comparison with and without OA ($N$=14, triplicates). The blue line represents results with OA ($R^2$ = 0.998, $y = 2.2x^{0.36}$), and the gray line represents results without OA ($R^2$ = 0.986, $y = 1.6x^{0.35}$). (d) Comparison of test spot intensities for the spiked samples within the 99$^{th}$ percentile range of cTnI concentration. (e) cTnI clinical sample test results and the impact of data-processing algorithms to exclude invalid results and improve repeatability; (i) before and (ii) after the assay validation and data refinement ($N_{initial}$ = 62, duplicated). Ground truth values of the samples with cTnI levels < 40 pg/mL are assigned to 39 pg/mL due to the clinical analyzer's cut-off level (40 pg/mL).

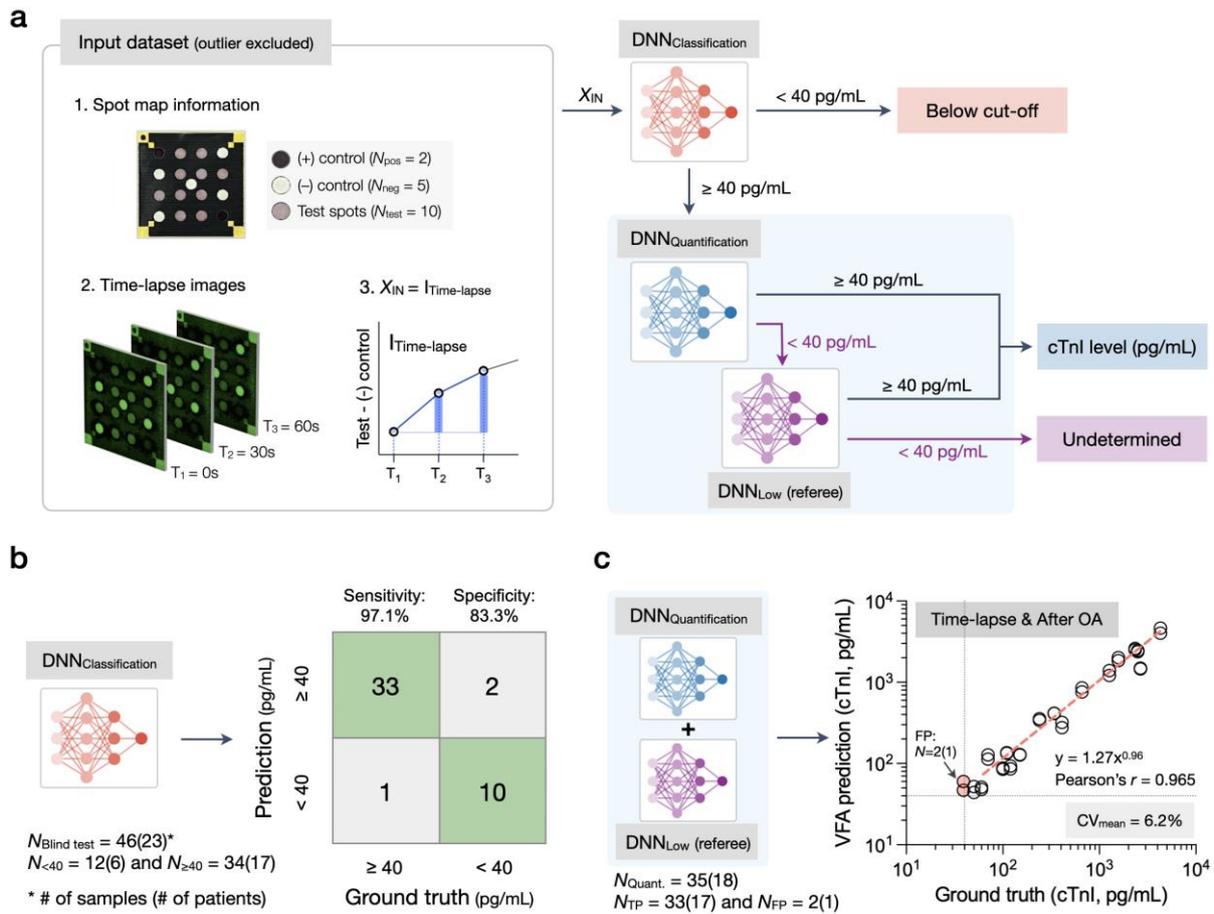

**Fig. 4.** Neural network-based analysis of cTnI concentrations in clinical serum samples. (a) Neural network-based signal processing pipeline. The input to the neural networks comprises a time-lapse signal calculated from three time-lapse images. The neural network-based processing pipeline consists of 1 classification ($DNN_{Classification}$) and 2 quantification ($DNN_{Quantification}$ and $DNN_{Low}$) neural networks that collaborate with each other. If a sample is blindly classified by the $DNN_{Classification}$ model as < 40 pg/mL (below the cut-off concentration), that is our final decision, and the quantification neural network models are not used in that case. Samples that are blindly classified as ≥ 40 pg/mL by $DNN_{Classification}$ are then processed by $DNN_{Quantification}$ for quantitative analysis. Blind inference results of $DNN_{Quantification}$ are used as our final concentration measurements if and only if the inference result is ≥ 40 pg/mL, in agreement with the $DNN_{Classification}$ blind inference. The samples predicted as < 40 pg/mL by $DNN_{Quantification}$ are further processed by $DNN_{Low}$ since this indicates a disagreement between the blind inference results of $DNN_{Classification}$ and $DNN_{Quantification}$. If the cTnI concentration predicted by $DNN_{Low}$ is still < 40 pg/mL the sample is labeled as "undetermined" as the disagreement with $DNN_{Classification}$ is further confirmed; otherwise, the prediction from $DNN_{Low}$ is used as the final concentration measurement. (b) Classification results from the optimized classification model ($DNN_{Classification}$) for 46 serum samples from 23 patients used in the blind testing set. The 40 pg/mL threshold concentration is due to the cut-off level of the benchtop clinical instrument used to obtain ground truth concentration measurements. (c) cTnI quantification results from the optimized quantification models ($DNN_{Quantification}$ and $DNN_{Low}$) for 35 serum samples from 18 patients classified into ≥ 40 pg/mL range by $DNN_{Classification}$ blind inference.

**Table 1.** Comparison of the presented hs-VFA platform against existing methods, including commercially available tests used for cTnI testing.

| Ref. | Platform (Category) | Sensing modality | Sample type (Volume) | LoD [pg/mL] | Assay range [ng/mL] | [a]Covers clinically relevant range (0.01 – 100 ng/mL) | Precision; CV(%) at 99th percentile level | Assay time | Types of reader device | Reader specs Size (L × W × D, cm) | Reader specs Weight (kg) |
|---|---|---|---|---|---|---|---|---|---|---|---|
| 22,23 | Abbott i-STAT (Commercial) | Electrochemical | Whole blood, plasma (17 µL) | 20 | ~ 50 | Insufficient (at low and high ranges) | 16.5 | 10 min | Portable | 20.9× 6.4 × 5.2 | 0.65 |
| 22,24 | Quidel/Alere Triage True (Commercial) | Fluorescence | Whole blood, plasma (175 µL) | 1.5 | [b]N/C | Indeterminate (N/C in assay range) | 6.4 | < 20 min | Benchtop & Portable | 22.5 × 19 × 7 | 0.68 |
| 22,25 | Siemens ATELLICA VTLi (Commercial) | Optical magnetic | Whole blood, plasma, capillary blood (30 – 100 µL) | 1.24 | 0.00124 – 1.25 | Insufficient (at high range) | 6.7 | 8 min | Portable | 25 × 8.5 × 5.2 | 0.78 |
| 26,27 | Philips Minicare (Commercial) | Optomagnetic | Whole blood, plasma (30 µL) | 17 | 0.018 – 7 | Insufficient (at low and high ranges) | 18.6 | < 10 min | Portable | N/C | N/C |
| 22,28 | Response Biomedical RAMP (Commercial) | Fluorescence | Whole blood (75 µL) | 30 | 0.03 – 32 | Insufficient (at low and high ranges) | 20 | 19 min | Benchtop & Portable | 25 × 25 × 15 | 2 |
| 22,29 | LSI Medicine PATHFAST (Commercial) | Chemiluminescence | Whole blood, plasma (100 µL) | 7 | 0.019 – 50 | Insufficient (at low and high ranges) | 6.1 | < 17 min | Benchtop | 75 × 57 × 51 | 33 |
| 22,30 | Radiometer AQT90 FLEX (Commercial) | Fluorescence | Whole blood, plasma (65 µL) | 9 | 0.01 – 25 | Insufficient (at high range) | 12.9 | < 19 min | Benchtop | 45 × 46 × 48 | 35 |
| 31 | Microarray (Research) | Near-infrared fluorescence | Serum (10 µL) | 15.2 | 0.016 – 1.2 | Insufficient (at low and high ranges) | < 15 | ~ 3 h | Benchtop | 28 × 46 × 37 | 15.5 |
| 32 | Lateral flow assay (Research) | Colorimetric | Serum (40 µL) | 9 | ~ 50 | Insufficient (at high range) | < 5 | > 25 min | Benchtop | N/C | N/C |
| 33 | Lateral flow assay (Research) | Chemiluminescence | Serum, plasma (50 µL) | 0.84 | 0.001 – 10 | Insufficient (at high range) | < 9 | 20 min | Benchtop | 30 × 60 × 96 | 32 |
| 36 | Field effect transistor (Research) | Electrochemical | Whole blood (< 10 µL) | < 80 | 0.1 – 24 | Insufficient (at low and high ranges) | N/C | 5 min | Benchtop | N/C | N/C |
| 37 | Field effect transistor (Research) | Electrochemical | Serum (N/C) | 0.3 | 0.001 – 1.4 | Insufficient (at high range) | N/C | < 20 min | Portable | N/C | N/C |
| 38 | Lateral flow assay (Research) | Surface-enhanced Raman scattering | Serum (Volume; N/C) | 20 | 0.01 – 100 | Sufficient | N/C | < 30 min | Benchtop | N/C | N/C |
| This work | Vertical flow assay (Research) | Colorimetric | Serum (50 µL) | 0.2 | 0.001 – 100 | Sufficient | < 5 | < 15 min | Portable | 15.5 × 10 × 5.5 | 0.46 |

[a] The article no.18 in the reference list establishes the clinically relevant range.
[b] "N/C" denotes "not commented" on the corresponding reference or the manufacturer's website.